\documentclass[floatfix,showpacs,prl,aps]{revtex4}
\usepackage[applemac]{inputenc}
\usepackage[T1]{fontenc} 
\pdfoutput=1
\usepackage{float}
\usepackage{amssymb}
\usepackage{enumerate}
\usepackage{braket}
 \usepackage{amsmath} 
\usepackage{subfigure}
\usepackage{lipsum}
\usepackage{amsfonts} 
\usepackage{amssymb, mathrsfs}
\usepackage{amsmath}
\usepackage{bm}
\usepackage[pdftex]{graphicx}
\usepackage{footnote}
\usepackage[colorlinks]{hyperref}
\usepackage{titlesec}
\usepackage{graphicx} 

\def\beq{\begin{equation}}
\def\eeq{\end{equation}}
\def\bsp{\begin{split}}
\def\esp{\end{split}}
\def\bea{\begin{eqnarray}}
\def\eea{\end{eqnarray}}
\def\ba{\begin{array}}
\def\ea{\end{array}}

\def\haf#1{{{#1}\over 2}}

\def\lb{\left(}
\def\rb{\right)}

\def\l.{\left.}
\def\r.{\right.}

\def\ie{{\it i.e. }}

\def\part{\partial}
\def\tfrac#1#2{{\textstyle{#1\over #2}}}
\def\half{\tfrac{1}{2}}

\newcounter{subsections}

\titleformat{\subsection}
  {}
  {0pt}
  {}                    
  {\bfseries\addtocounter{subsections}{1}\thesubsections. } 

\begin{document}

\preprint{UdeM-GPP-TH-19-274}
\preprint{arXiv:1910.01774}
\title{Stable, thin wall, negative mass bubbles in de Sitter space-time.}
\author{Matthew C. Johnson$^{3,5}$}
\email{mjohnson@perimeterinstitute.ca}
\author{M. B. Paranjape$^{1,2,3}$} 
\email{paranj@lps.umontreal.ca}
\author{Antoine Savard$^{1}$} 
\email{antoine.savard@umontreal.ca}
\author{Natalia Tapia-Arellano$^{1,4}$} 
\email{natalia.tapiaa@usach.cl}

\affiliation{$^{1}$Groupe de physique des particules, D\'epartement de physique,
Universit\'e de Montr\'eal,
C.P. 6128, succ. centre-ville, Montr\'eal, 
Qu\'ebec, Canada, H3C 3J7 }
\affiliation{$^{2}$Centre de recherche mathématiques,
Universit\'e de Montr\'eal}
\affiliation{$^{3}$Perimeter Institute for Theoretical Physics, 31 Caroline Street North, Waterloo, Ontario, Canada N2L 2Y5 }
\affiliation{$^4$ Departamento de F\'\i sica, Universidad de Santiago de Chile, Avenida Ecuador 3493, Estación Central, Santiago, Chile 9170124.}
\affiliation{$^{5}$Department of Physics and Astronomy,  128 Petrie Science and Engineering Building, York University
4700 Keele Street, Toronto, Ontario, Canada, M3J 1P3 }
\begin{abstract}

\section{Abstract}  
Negative mass makes perfect physical sense as long as the dominant energy condition is satisfied by the corresponding energy-momentum tensor. Heretofore, only {\it configurations} of negative mass had been found \cite{Belletete:2013nqa,Mbarek:2014ppa}, the analysis did not address stability or dynamics.  In this paper, we analyze both of these criteria.  We demonstrate the existence of {\it stable}, static, negative mass bubbles in an asymptotically de Sitter space-time.  The bubbles are solutions of the Einstein equations and correspond to an interior region of space-time containing a specific mass distribution, separated by a thin wall from the exact, negative mass Schwarzschild-de Sitter space-time in the exterior.  We apply the Israel junction conditions at the wall.  For the case of an interior corresponding simply to de Sitter space-time with a different cosmological constant from the outside space-time, separated by a thin wall with energy density that is independent of the radius, we find static but unstable solutions which  satisfy the dominant energy condition everywhere.  The bubbles can collapse through spherically symmetric configurations to the exact, singular, negative mass Schwarzschild-de Sitter solution.  Interestingly, this provides a counter-example of the cosmic censorship hypothesis.  Alternatively, the junction conditions can be used to give rise to an interior mass distribution that depends on the potential for the radius of the wall.  We show that for no choice of the potential, for positive energy density on the wall that is independent of the radius,  can we get a solution that is non-singular at the origin.  However, if we allow the energy density on the wall to depend on the radius of the bubble, we can find {\it stable}, static, non-singular solutions of negative mass which everywhere satisfy the dominant energy condition.  
\end{abstract}


\maketitle


\subsection{Introduction}
The Schwarzschild metric is a solution of the vacuum Einstein equations with one parameter, the mass.  It is a solution of the Einstein equations for any value of the mass, including negative mass.  However it is a singular solution, the singularity residing at the origin of the coordinate system.  The singularity means that in some sense the solution actually contains a source, a singular source located at the position of the singularity.  The positive mass singularity is hidden behind an event horizon while the negative mass singularity is naked.  Smoothing out the singularity corresponds to adding an energy-momentum source to the space-time.  The smoothed metric satisfies Einstein equations with this energy-momentum as the source.   The negative mass singularity cannot be smoothed out with a source that could correspond to physically sensible energy-momentum.  Physically sensible energy-momentum is taken to mean that at any point, the flow of the energy-momentum remains inside the future directed light-cone from that point.  Such energy-momentum satisfies the dominant energy condition, which means, technically, for any future directed time-like or light-like vector $u$ :
\beq
T^{0 \nu}u_{\nu} \geq 0\quad{\rm and }\quad T^{\mu \nu}u_{\nu} T_{\mu \alpha}u^{\alpha}  \geq 0\label{dec}
\eeq 
If the dominant energy condition is satisfied, then one can prove the positive energy theorem  \cite{Schon:1979rg,Schon:1981vd,Witten:1981mf} which implies that the ADM mass \cite{Arnowitt:1962hi} must be positive, denying the possibility of negative mass.  The positive energy theorem requires an asymptotically flat space-time or asymptotically anti-de Sitter space-time.  Negative mass solutions have already been found in anti-de Sitter space-time \cite{Mann:1997jb,Smith:1997wx}, however they cannot satisfy the dominant energy condition.  But in asymptotically de Sitter space-time, the positive energy theorem does not hold, and it is here that one could imagine that physically reasonable, non-singular negative mass solutions could exist.   The first example of such a possibility of negative mass was found in the article \cite{Belletete:2013nqa}.  In this work a simple mathematical deformation of the negative mass Schwarzschild-de Sitter metric
\beq
ds^2=\left(1-\frac{2(-M+\Lambda r^3/6)}{r}\right)dt^2-\left(1-\frac{2(-M+\Lambda r^3/6)}{r}\right)^{-1}dr^2-r^2d\Omega^2\label{sds}
\eeq
was given.  By changing $(-M+\Lambda r^3/6)\to m(r)$ and imposing that $m(r)\to 0$ for $r\to 0$  but $m(r)=-M+\Lambda r^3/6$ for sufficiently large $r$ yields an asymptotic, negative mass Schwarzschild-de Sitter space-time and the singularity at the origin is smoothed out.  When this metric is inserted into the Einstein tensor, the result no longer vanishes, and the result is taken to be equal to the required energy-momentum tensor to satisfy Einstein's equations.  It was shown in \cite{Belletete:2013nqa} that it was possible to choose the deformation so that the resulting energy-momentum tensor satisfies the dominant energy condition everywhere.  This work established the possibility that non-singular negative mass configurations could exist that arise out of physically reasonable energy-momentum.   The notion that the negative mass so obtained is spurious because it is defined only relative to the background, is groundless.  There is no intrinsic notion of mass in asymptotic de Sitter space-time, \cite{Ashtekar:2014zfa,Ashtekar:2015lla,Ashtekar:2015lxa}.  Even positive mass in such a space-time is only defined relative to the background and therefore positive or negative mass configurations are equally valid.

In a subsequent article, \cite{Mbarek:2014ppa} it was shown that with energy and momentum corresponding to that of an ideal fluid, there exist bubble like configurations with the exterior space-time given exactly by the negative mass Schwarzschild-de Sitter space-time.  However also in \cite{Mbarek:2014ppa}, no dynamics were accorded to the ideal fluid, no equation of state was imposed and therefore the bubbles that were found were again just configurations and not solutions of a dynamical theory.  The energy-momentum of an ideal fluid, in the comoving coordinate system, is characterized by two functions, the pressure and the density, and the dominant energy condition corresponds to:
\beq
\rho(r)\ge 0 \quad\quad \rho(r)\ge |p(r)|\label{dec1} 
\eeq
Einstein's equations are under-determined giving three equations for four fields assuming spherical symmetry, the pressure, the density and the coefficient fields of $dt^2$ and $dr^2$ in the metric.  Usually an equation of state relating $\rho$ to $p$ is specified, giving rise to a deterministic system.  Instead of providing the equation of state, in \cite{Mbarek:2014ppa}, the coefficient field of $dr^2$ in the metric was simply specified.  It was smoothly and explicitly  deformed inside a radius $y$ till $r=0$ from its value in the negative mass Schwarzschild-de Sitter metric outside, in a manner that eliminated the singularity at the origin.   Einstein's dynamical equations were solved (numerically) for the coefficient field of $dt^2$ in the metric, and for the density and pressure.  It was observed that the dominant energy condition Eqn. \eqref{dec1} was satisfied.  Thus it was shown that perfectly physical matter, that corresponding to an ideal fluid,  could in principle organize itself to correspond to localized regions of negative mass.   

However, this work was still unsatisfactory, stability of the solution was not addressed.   It is still desirable to find a dynamical system in which actual self-consistent, soliton-like solutions of the dynamical matter/Einstein equations would exist and hence give rise to stable, non-singular, static solutions which correspond to localized regions of negative mass.  In this article we show how to obtain dynamically {\it static}, non-singular solutions of negative mass which satisfy the dominant energy condition everywhere.  The solutions are comprised of and inner and an outer space-time separated by a thin wall.  We  obtain stable solutions if the energy density on the wall is non constant \ie is a (rather simple) function of the radius or if it breaks the dominant energy condition.  

\subsection{The setup}
We will construct our solutions assuming a spherical geometry and using Schwarzschild coordinates.  The solution will correspond to the exact negative mass Schwarzschild-de Sitter geometry given by Eqn. \eqref{sds} outside, with $\Lambda\rightarrow\Lambda_{e}$, separated by a thin wall from an inside geometry.    The conservation of energy and momentum across the wall is obtained by imposing the Israel junction conditions  \cite{Israel:1966rt}.   A clear exposition of the application of the Israel junction conditions is given in \cite{Visser:2003ge}.  The wall is characterized by two parameters, the energy density per unit area $\sigma$, and the surface tension $\vartheta$ (defined  so that $\vartheta$ is positive if the surface wants to contract and negative if the surface wants to expand).  Then the stress-energy tensor of the wall will have the form (in an orthonormal system of coordinates)
\beq
S_{\hat a\hat b}={\rm diag.}\,\, (\sigma,-\vartheta,-\vartheta).
\eeq
Imposing the Israel junction conditions will permit us to find the necessary inside geometry that will give rise to non-singular, stable solutions.  

The interior mass function is taken to be $m_-(r)$ which is not specified while the exterior mass function is taken to be explicitly
\beq
m_+(r)=-M+\frac{\Lambda r^3}{6}
\eeq
corresponding to an exact negative mass Schwarzschild-de Sitter space-time with mass $-M$ cosmological constant $\Lambda$.  
The balance of energy-momentum flux through the interface gives rise to the junction condition in our case:
\beq
\left(1-\frac{2m_-(r)}{r}+ \dot{r}^{2}\right)^{1/2}-
\left(1+\frac{2M}{r} -\frac{\Lambda r^{2} }{3} + \dot{r}^{2} \right)^{1/2} = 4 \pi \sigma r \label{ijc}
\eeq
and 
\beq
\frac{\left(1-\frac{m_-(r)}{r}-m_-'(r)+\dot{r}^2+r\ddot{r}\right)}{\left(1-\frac{2m_-(r)}{r}+ \dot{r}^{2}\right)^{1/2}} - \frac{\left(1+\frac{M}{r}-\frac{2\Lambda r^2}{3}+\dot{r}^2+r\ddot{r}\right)}{\left(1+\frac{2M}{r} -\frac{\Lambda r^{2} }{3} + \dot{r}^{2} \right)^{1/2}} = 8 \pi \vartheta r \label{ijc2}
\eeq
where $\Lambda$ is the vacuum energy on the outside $-M$ is the value of the mass of the configuration as viewed from the outside and $m_-(r)$ is the mass function inside.  Writing this Eqn.\eqref{ijc} as simply
\beq
\lb a +\dot r^2\rb^{1/2}-\lb b +\dot r^2\rb^{1/2}=c^{1/2}\label{10}
\eeq
with obvious expressions for $a,b,c$, we can easily solve for $\dot r^2$ by squaring both sides, reorganizing to put the square roots to one side and squaring again, which yields
\beq
\lb a+b-c\rb^2 -4 ab= 4c \dot r^2
\eeq
\ie
\beq
\dot r^2=\frac{(a-b)^2}{4c}-\frac{(a+b)}{2}+\frac{c}{4}\label{rdot}
\eeq
The LHS of Eqn.\eqref{rdot} can be thought of as (twice) the negative of the potential, $V(r)$ and the dynamics  corresponds to motion in this potential with vanishing total energy.  Thus Eqn.\eqref{rdot} can be written as a dynamical equation for $r$
\beq
 \frac{\dot r^2}{2} +V(r)=E=0.
\eeq  
where 
\beq
V(r)=-\half \lb \frac{(a-b)^2}{4c}-\frac{(a+b)}{2}+\frac{c}{4}\rb\label{12}.
\eeq
Replacing for $a$, $b$ and $c$ we get
\beq
V(r)=-\half\lb\frac{\lb m_-(r)+M-\frac{\Lambda r^3}{6}\rb^2}{16\pi^2\sigma^2 r^4}-\lb 1+\frac{(m_-(r)-M)}{r}+\frac{\Lambda r^2}{6}\rb+4\pi^2\sigma^2 r^2\rb\label{V}.
\eeq
A static stable  solution will arise with a potential that admits a radius $r_0$ such that 
\begin{equation}
V(r_0)=0, \quad V'(r_0)=0,\quad V''(r_0) >0.
\end{equation}
Alternatively, we can use Eqn.\eqref{V}  to solve for $m_-(r)$ in terms of the potential $V(r)$ or equivalently Eqn.\eqref{12} to solve for $a$.  We have
\beq
(a-b)^2 -(a-b)2c +c^2 -4bc +8V(r)c=0
\eeq
Thus
\beq
a-b=c\pm\sqrt{c^2-(c^2-4bc +8V(r)c)}=c\pm 2\sqrt{c(b-2V(r))}
\eeq
or
\beq
a=c+b\pm 2\sqrt{c(b-2V(r))}.
\eeq
Solving for $a$ in Eqn.\eqref{10} when $\dot r=0$ and the $V(r)=0$ shows that we must take the + sign here.  Replacing for  $a$, $b$ and $c$ we get
\beq
1-\frac{2m_-(r)}{r}=\lb 4 \pi \sigma r\rb^2 +1+\frac{2M}{r} -\frac{\Lambda r^{2} }{3}+2\sqrt{\lb 4 \pi \sigma r\rb^2\lb 1+\frac{2M}{r} -\frac{\Lambda r^{2} }{3}-2V(r)\rb}
\eeq
or
\beq
m_-(r)=-8 \pi^2 \sigma^2 r^3 -M +\frac{\Lambda r^{3} }{6}-4\pi \sigma r^2\sqrt{ 1+\frac{2M}{r} -\frac{\Lambda r^{2} }{3}-2V(r)}.\label{19}
\eeq
One can see that for no choice of the potential, $m_-(0) = 0$ for constant $\sigma$.  This means that it is not possible to choose the potential so that the solution will be non-singular at the origin. 
\subsection{de Sitter bubbles}
We begin by considering the case where the bubble interior is pure de Sitter with a different cosmological constant than the ambient de Sitter space-time, such that $m_-(r)=\frac{\Lambda_i r^3}{6}$. Vacuum bubbles of this type have been studied in a variety of contexts~\cite{Berezin:1982ur,Blau:1986cw,Farhi:1986ty,Berezin:1987bc,Farhi:1989yr,Aurilia:1989sb,Mazur:2004fk,Aguirre:2005xs,Aguirre:2005nt}. Here, we extend these previous analyses to negative mass in the exterior space-time as was analyzed by Barnaveli and Gogberashvili \cite{Barnaveli:1996pn,Barnaveli:1995ep,Barnaveli1994}. These authors did find the static, unstable negative mass solutions that we will expose in this section, however they did not find the stable solutions that we will reveal in subsection (5).  We can write
\beq
V(r)=-\frac{\alpha}{r^4}+\frac{\beta}{r}-\gamma\, r^2+\frac{1}{2}\label{radius}
\eeq
where
\begin{eqnarray}
\alpha &=& \frac{M^2 }{32 \pi^{2} \sigma^{2}} , \\
\beta &=& \frac{M \left(\Lambda-\Lambda_i \right)}{96 \pi^{2} \sigma^{2}}    + \frac{M}{2} , \\
\gamma &=& \frac{\left( \Lambda_i - \Lambda \right)^2}{1152 \pi^2 \sigma^{2}} + \frac{\lb\Lambda_{i}+\Lambda\rb}{12} + 2 \pi ^2 \sigma ^2\label{21}
\end{eqnarray}
where $\sigma$ is taken to be a constant independent of the bubble radius.  
The coefficients $\alpha,\gamma$ are positive while $\beta$ can have any sign. But for large and small $r$, the $\beta/r$ term is unimportant.   The potential $V(r)$ obviously descends to $-\infty$ in both limits, $r\to 0$ and $r\to\infty$, and it is easy to show that it has exactly one maximum in between.  The derivative of the potential, set equal to zero gives
\beq
2V'(r)=\frac{4\alpha}{r^5}-\frac{\beta}{r^2}-2\gamma\, r=\frac{{4\alpha}-{\beta}{r^3}-2\gamma\, r^6}{r^5}=0.\label{vp}
\eeq
The numerator is a simple quadratic in $r^3$ with solutions
\beq
r_\pm^3=\frac{\beta\pm\sqrt{\beta^2+32\alpha\gamma}}{-4\gamma}.
\eeq
The two roots are positive and negative, the cube root maintains the sign, and we discard the negative root.  Thus we find one positive root where the derivative of the potential vanishes.  The positive root is always
\beq
r_0=\lb\frac{\sqrt{\beta^2+32\alpha\gamma}-\beta}{4\gamma}\rb^{1/3}.\label{pr}
\eeq
For this value of the radius to give a static solution requires that the potential vanish
\beq
V(r_0)=0\label{r0}
\eeq
which can be simply arranged by choosing the parameters $M,\Lambda, \Lambda_i$ and $\sigma$.  
Due to the asymptotic behaviour of the potential, this extremum must be a maximum.  
\begin{figure}[ht]
\centerline{\includegraphics[width=0.5\linewidth]{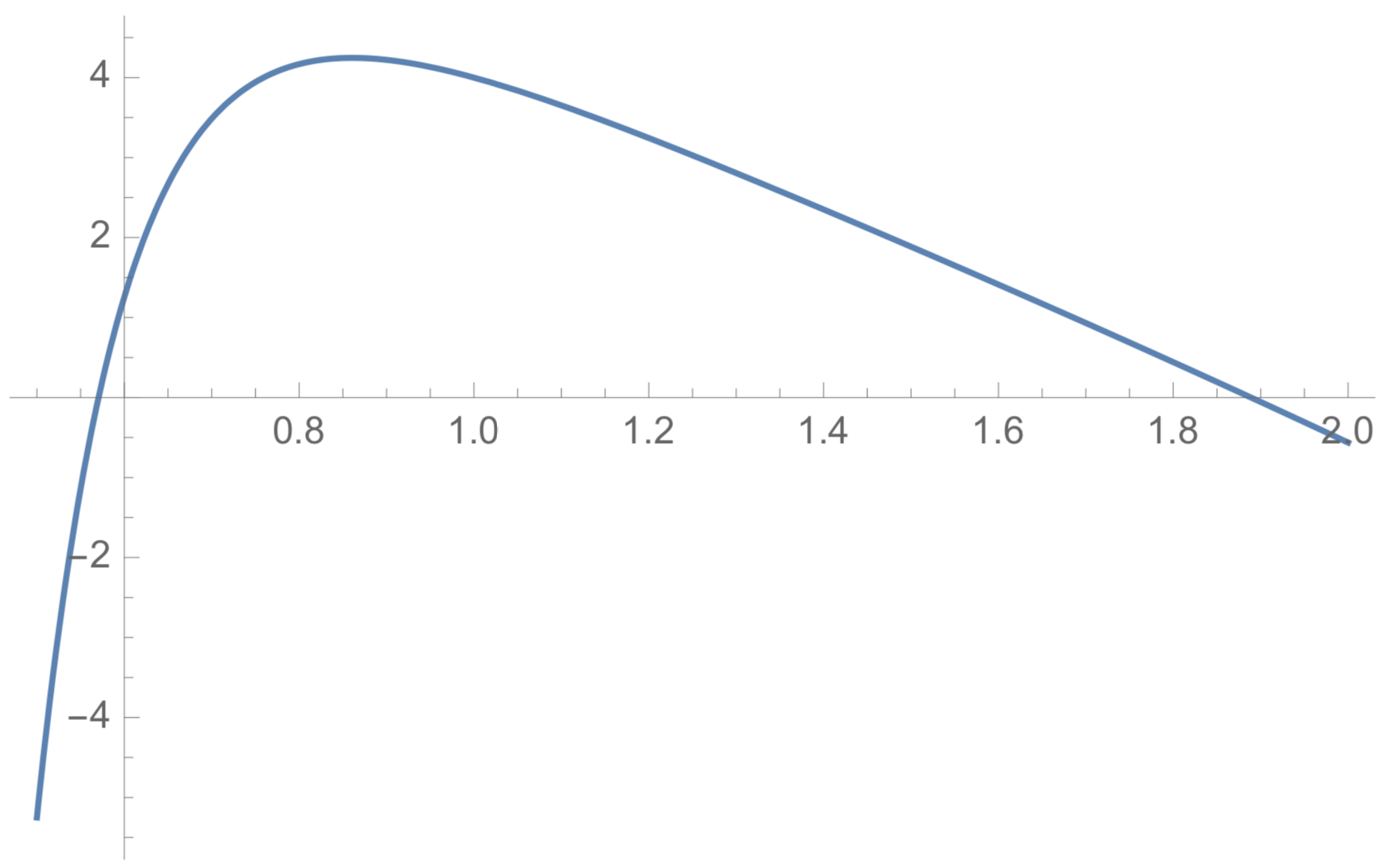}} \caption{\label{fig0} (colour online) The potential $V(r)$ plotted as a function of $r$ (in units of mass) for generic values of $\alpha,\beta,\gamma$}
\end{figure}
Thus it is obvious that one can have an unstable, negative mass bubble when the parameters are chosen so that the top of the potential has a double root at its maximum. This solution is the negative mass analog to the unstable solutions of Refs.~\cite{Garriga:2004nm,Aguirre:2005nt,Aguirre:2005xs}.

\begin{figure}[ht]
\centerline{\includegraphics[width=0.5\linewidth]{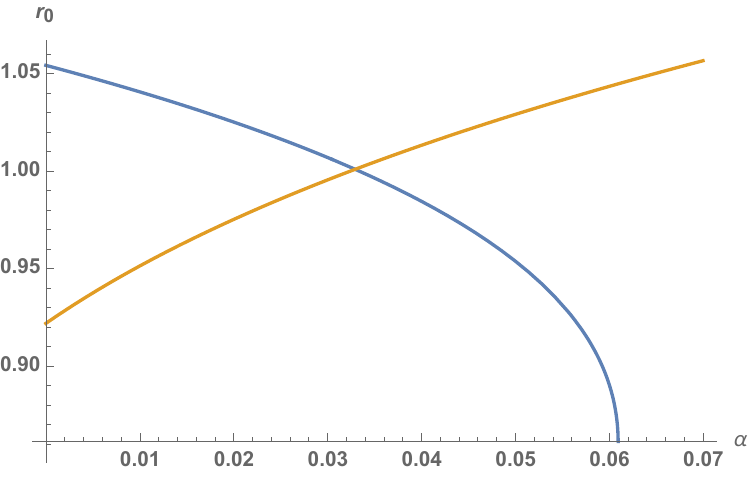}} \caption{\label{r_0} (colour online) The the curves for $r_0$ as a function of $\alpha\approx .0165$, with a $\beta=-.235 $ and $\gamma=.3$, with crossing point giving the solution for $V(r_0)=0$ and  $V'(r_0)=0$ which occurs at about $r_0\approx 1$.}
\end{figure}

We can find such a solution by solving  the equation for $V(r_0)=0$ with $r_0$  as given in Eqn.\eqref{pr} (such that $V'(r_0)=0$).  From Eqn.\eqref{radius} we can write $V(r_0)=0$ implies
\beq
-\alpha +\beta r_0^3-\gamma r_0^6 +\frac{r_0^4}{2}=0.
\eeq
Then replacing from $V'(r_0)=0$ that gives $\beta r_0^3=4\alpha-2\gamma r_0^6$ from Eqn.\eqref{vp} we get
\beq
6\alpha-6\gamma r_0^6+r_0^4=0,\label{cubic}
\eeq
which is an easily, analytically solved cubic in the $r_0^2$.  One can easily solve for the root analytically, and then plot the two solutions for $r_0$ from Eqn.\eqref{pr} and from the solution of the cubic Eqn.\eqref{cubic} giving the curves in Fig.\eqref{r_0}.  The crossing of the curves gives the value of $r_0$ for which $V(r_0)=0$ and  $V'(r_0)=0$.
\vfill\break

\begin{figure}[h]
\centerline{\includegraphics[width=0.5\linewidth]{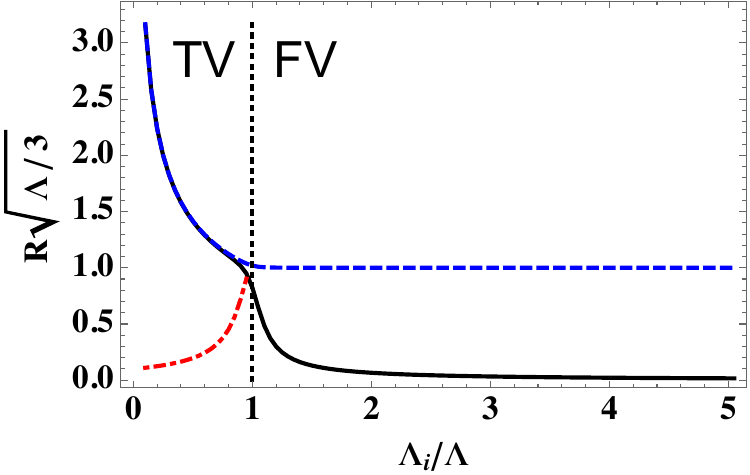}} \caption{\label{fig:dsbubble} (colour online) The radius of unstable negative mass vacuum bubbles (solid black) as a function of $\Lambda_i / \Lambda$. The regions corresponding to True Vacuum (TV) bubbles $\Lambda_i / \Lambda < 1$ and False Vacuum (FV) bubbles $\Lambda_i / \Lambda > 1$ are indicated. We overplot the radius of the cosmological horizon (blue dashed) and the zero mass Coleman-de Luccia true vacuum bubbles (red dot-dashed).}
\end{figure}

Static but unstable negative mass solutions exist for a variety of parameters, including for both true vacuum (e.g. $\Lambda > \Lambda_i$) or false vacuum (e.g. $\Lambda < \Lambda_i$) bubbles. Assuming fixed $\sigma$, in Fig.~\ref{fig:dsbubble} we plot the radius of the unstable solution as a function of the ratio $\Lambda_i / \Lambda$. The mass of the static unstable solutions increases as the ratio $\Lambda_i / \Lambda$ decreases. The radius of the cosmological horizon (blue dashed line) in negative mass Schwarzschild de Sitter grows with increasing magnitude of the mass parameter~\footnote{This is a novel property of negative mass Schwarzschild-de Sitter, implying that the entropy associated with the cosmological horizon can grow without bound. This can be contrasted with positive mass Schwarzschild-de Sitter, where the total entropy associated with the cosmological and black hole horizons is bounded by the magnitude of $\Lambda$.}, and unstable true vacuum bubbles track this growth, remaining just inside the cosmological horizon. False vacuum bubbles on the other hand are parametrically smaller than the cosmological horizon, decreasing in size as the ratio $\Lambda_i / \Lambda$ increases. We can contrast these properties with the zero mass Coleman-de Luccia (CDL) true/false vacuum bubbles~\cite{Callan:1977pt,Coleman:1977py,Coleman:1980aw} or positive mass true/false vacuum bubbles~\cite{Berezin:1982ur,Blau:1986cw,Farhi:1986ty,Berezin:1987bc,Farhi:1989yr,Aurilia:1989sb,Mazur:2004fk,Aguirre:2005xs,Aguirre:2005nt}. CDL true vacuum bubbles are always smaller than the cosmological horizon, as shown in Fig.~\ref{fig:dsbubble} (red dot-dashed line), and increase in size with the ratio $\Lambda_i / \Lambda$; this trend is opposite that of negative mass unstable true vacuum bubbles. CDL false vacuum bubbles are always larger than the cosmological horizon. Qualitatively similar behavior is found for positive mass bubbles which do not collapse to a singularity. We can therefore see that negative mass bubbles are qualitatively very different than their positive mass counterparts. 

For other generic values of the parameters, there may be no turning points when the potential never crosses zero. In this case, initially expanding bubbles will continue to expand to infinite size, while initially collapsing bubbles will shrink away to zero size. When the parameters give a potential that does cross zero, the solutions split into two classes, those of radius greater than the larger zero crossing, which collapse to the minimum size given by the radius of the zero crossing and then bounce back to expand to infinite radius, and those of radius smaller than the smaller zero crossing which will expand to a maximum size given by the radius of the smaller zero crossing, and then shrink back down to zero size. This is all in exact analogy with the qualitative types of solutions that exist for positive mass bubbles~\cite{Berezin:1982ur,Blau:1986cw,Farhi:1986ty,Berezin:1987bc,Farhi:1989yr,Aurilia:1989sb,Aguirre:2005xs,Aguirre:2005nt}. 

For static unstable vacuum bubbles, the dominant energy condition reduces to $\sigma \geq |\vartheta|$. Conservation of stress energy implies $\sigma=\vartheta$, and it is therefore clear that our mass distributions satisfy the dominant energy condition. The other solutions described above also respect the dominant energy condition. Hence we can start with an initial physical mass distribution, that satisfies the dominant energy condition everywhere, but that in the latter case shrinks down to the singular solution corresponding to the exact negative mass Schwarzschild-de Sitter space-time which has a naked singularity.  Thus we have shown there exists perfectly physical initial data, that satisfies the dominant energy condition, which collapses to a naked singularity, albeit of negative mass. This is a counterexample to the cosmic censorship hypothesis. \cite{Penrose:1969pc}\footnote{We thank E. Wilson-Ewing for pointing this out to us.} Of course, we have not considered non-spherical perturbations. Because such perturbations grow in a collapsing bubble~\cite{Adams:1989su,Aguirre:2005xs}, the system might evade the singular negative mass solution by fragmentation or gravitational radiation under non-spherical perturbations.

\subsection{Stable negative mass solutions for constant $\sigma\le 0$.}
We take the  interior mass function given in Eqn.\eqref{19} 
and rescale the parameters to have only dimensionless parameters which allows us to factor out one power of M: $\Lambda=\frac{ 3\breve\Lambda}{M^2}$, $=\frac{\breve\sigma}{4\pi\sigma M}$,  $r=M\breve r$ such that the Eqn.\eqref{19} can be rewritten
\beq
\breve m_-(\breve r)=M\left(-1+\frac{1}{2}\left(\breve\Lambda-\breve\sigma^2\right)\breve r^3-\breve\sigma\breve r^2\sqrt{1-2\breve V(\breve r)+\frac{2}{\breve r}-\breve\Lambda\breve r^2}\,\right).\label{27}
\eeq
We will drop the breves in the subsequent analysis.  We have already observed that for any choice of the potential, it is impossible to remove the singularity at $r=0$ in the metric which comes in the form $\frac{m_-(r)}{r}$, the term $-1$ in Eqn.\eqref{27} cannot be cancelled unless we take $\sigma$ is negative, which we will briefly examine below.  Such a detour affords a stable solution, but evidently one that is not physical, the energy density on the wall must be negative and the dominant energy condition is not satisfied.  Then Eqn.\eqref{27} becomes, with $\sigma\to-\sigma$
\beq
 m_-( r)=M\left(-1+\frac{1}{2}\left(\Lambda-\sigma^2\right) r^3+\sigma r^2\sqrt{1-2 V( r)+\frac{2}{ r}-\Lambda r^2}\,\right)
\eeq
where now $\sigma$ is taken positive.
Making the further substitution
\beq
V(r)=-\frac{1}{2\sigma^2 r^4}\left(1+\tilde V(r)\right)\label{vtilde},
\eeq
we find
\beq
m_-( r) = M\left( -1 +\frac{1}{2} \left( \Lambda -\sigma^2\right) r^3 +\sqrt{1+\tilde V( r)+\sigma^2 r^4 +2\sigma^2 r^3 -\Lambda\sigma^2 r^6}\,\right) \label{mi1}
\eeq
and we see that as $r\to 0$ we also have  $m_-(r)\to 0$ as long as $\tilde V(r)$ also vanishes at the origin, so that any potential singularity in the metric at the origin is evaded. 
With the additional definition
\beq
(1+U(r))^2= 1+\tilde V(r)+\sigma^2r^4 +2\sigma^2 r^3 -\Lambda\sigma^2 r^6
\eeq
we get the very simple expression
\beq
m_-(r)=M\lb \half \lb \Lambda -\sigma^2\rb r^3 +U(r)\rb  ,
\eeq
We note that the potential singularity in the metric at the origin due to the $-1$ term in Eqn.\eqref{mi1} has been neatly made to cancel.  
The dominant energy condition Eqn.\eqref{dec1} can be re-expressed in terms of the mass function \cite{Belletete:2013nqa} 
\begin{equation}
\frac{d}{dr}\left(\frac{m_-'(r)}{r^2} \right) \leq 0 
\quad{\rm and}\quad
\frac{d}{dr}\left(m_-'(r)r^2 \right) \geq 0 \label{dec12}
\end{equation}
then with this reparametrization,   we get 
\begin{equation}
\frac{d}{dr}\left(\frac{U'(r)}{r^2} \right) \leq 0 
\quad{\rm and}\quad
6 \lb \Lambda -\sigma^2\rb+\frac{d}{dr}\left(U'(r)r^2 \right) \geq 0 .
\end{equation}
The idea now is to choose the potential $\tilde V(r)$, and consequently $U(r)$, so that the three required conditions are satisfied: first that $\tilde V(r)$ has a double zero giving rise to a minimum at a given radius,  second to make sure that the dominant energy condition is satisfied at least in the bulk and third to ensure that $m_-(r)$ vanishes at the origin so that there is no singularity.   Such a minimum would actually give rise to an unstable maximum at the same radius for $V(r)$ as can be ascertained by looking at Eqn.\eqref{vtilde}.  However, it is easy to add small perturbations to $\tilde V(r)$ afterwards to convert the maximum into a local, stable minimum in $V(r)$ since around the maximum, which is also a double zero, the magnitude of $V(r)$ is arbitrarily small.   Sustained attempts to find a solution analytically in terms of a sixth order polynomial function for $U(r)$, which would have given rise to an analytical solution, were not successful.  It seems that monomials of all orders are required.   However, we have been able to find a numerical solution for a stable, nonsingular, negative mass bubble, which we expose below.  

We additionally rescale the radial variable so that we insist that the radius of the thin wall bubble is 1.  Thus replacing $r\to r/\rho $ but then also re-scaling all the coupling constants $\sigma\to \sigma\rho^{3/2}$, $\Lambda\to\Lambda\rho^{3}$ and $\tilde V(r)\to \tilde V(r/\rho)$ (which we will continue to call $\tilde V(r)$) simply inserts a factor of $1/\rho$ in the $r^4$ term under the square root in Eqn.\eqref{mi}
\beq
m_-(r)=M\lb -1 +\half \lb \Lambda -\sigma^2\rb r^3 +\sqrt{1+\tilde V(r)+\sigma^2 r^4/\rho +2\sigma^2 r^3 -\Lambda\sigma^2 r^6}\rb .\label{mi}
\eeq
however now, with the choice of $\rho$ equal to the putative bubble radius.

\begin{figure}[H]
\centerline{\includegraphics[width=0.5\linewidth]{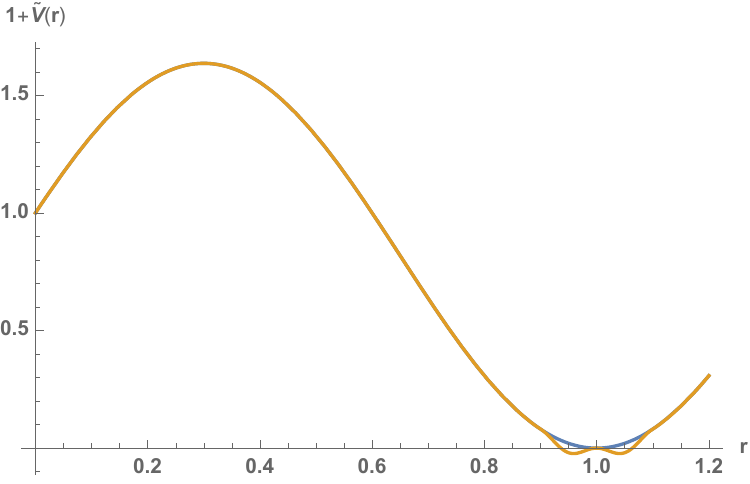}} \caption{ (colour online) The potential $1+\tilde V(r)$ and its perturbed version. As one can see, it only affects the potential in the vicinity of $r=1$. The parameters used are $\Lambda=0.1$, $M=1$, $\sigma=2$, $r_*=0.3$ and $\rho=0.05$.}\label{1pvtilde}
\end{figure}

\begin{figure}[H]
\centerline{\includegraphics[width=0.5\linewidth]{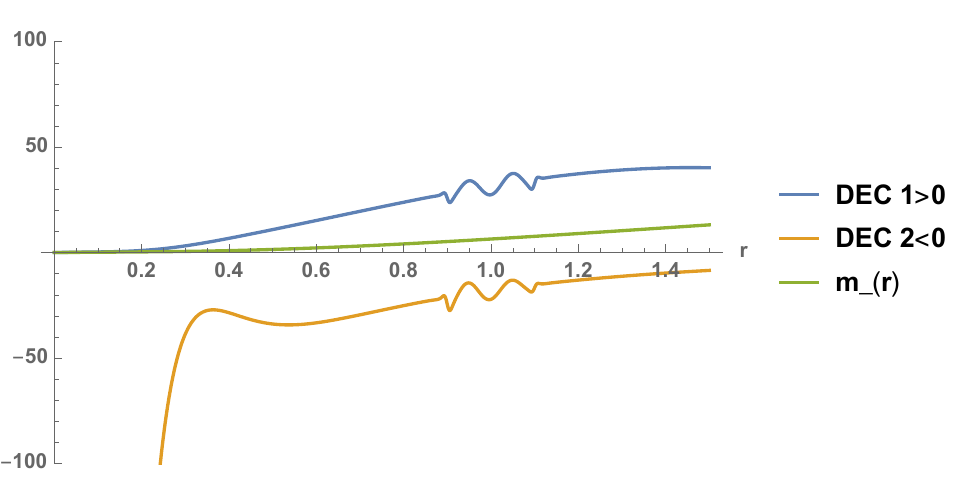}} \caption{ (colour online) The dominant energy conditions Eqn.\eqref{dec12} labelled here as (dec1) and (dec2) and the interior mass labelled Eqn.\eqref{mi} here as  $m_-$. The parameters used are $\Lambda=0.1$, $M=1$, $\sigma=2$, $r_*=0.3$ and $\rho=0.05$.}\label{dec12mm}
\end{figure}

A little numerical experimentation quickly shows that to satisfy the dominant energy condition, $1+\tilde V(r)$ must leave the origin at 1 with a positive slope. But then it can come down to have a smooth, double zero at $r=1$.  A possible choice, satisfying all the conditions is the simple function
\beq
1+\tilde V(r)=\cos^2\lb\frac{\pi(r-r_*)}{2(1-r_*)}\rb\sec^2\lb\frac{\pi r_*}{2(1- r_*)}\rb \label{eq1pvtilde}
\eeq
with $r_*$, in principle a free  parameter, works well for $r_*\approx 0.3$ and which is shown in Fig.\eqref{1pvtilde}.  
Then the dominant energy conditions and the interior mass functions are computed numerically as seen in Fig.\eqref{dec12mm}. A simple smooth deformation of $1+\tilde V(r)$ allows for a stable solution, giving rise to a local minimum at $r=1$ in the true potential as in Fig.\eqref{potp}.  The perturbation that was added is given by 
\beq
\delta\tilde V=\sin^2\lb10 r \pi\rb \lb0.1\tanh\lb100 (r - 0.9)\rb + 0.1\rb\lb 0.1\tanh\lb 100 (r - 1.1)\rb - 0.1\rb
\eeq
which when added to $\tilde V(r)$ gives the potential in Fig.\eqref{1pvtilde}.  We ensure (numerically) that the deformation does not disturb the dominant energy conditions, as can be seen in Fig.\eqref{dec12mm}.  This solution is stable, non-singular and satisfies the dominant energy condition in the bulk, however, the solution is still not satisfactory as the energy-momentum on the wall is not physical.  

\begin{figure}[H]
\centerline{\includegraphics[width=0.5\linewidth]{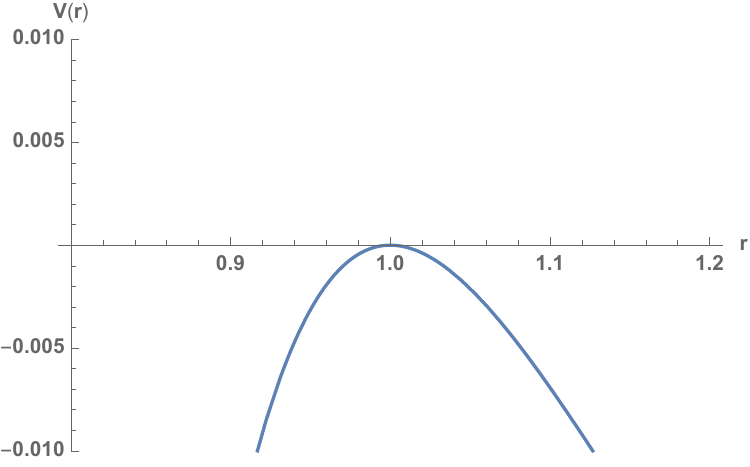}} \caption{ (colour online) The potential $V(r)=-\frac{1}{2\sigma^2 r^4}(1+\tilde V(r))$ with $1+\tilde V$ as in Eqn.\eqref{eq1pvtilde} without any perturbations. Note here that there is a zero at $r=1$. The parameters used are $\Lambda=0.1$, $M=1$, $\sigma=2$, $r_*=0.3$ and $\rho=0.05$.}\label{potsp}
\end{figure}

\begin{figure}[H]
\centerline{\includegraphics[width=0.5\linewidth]{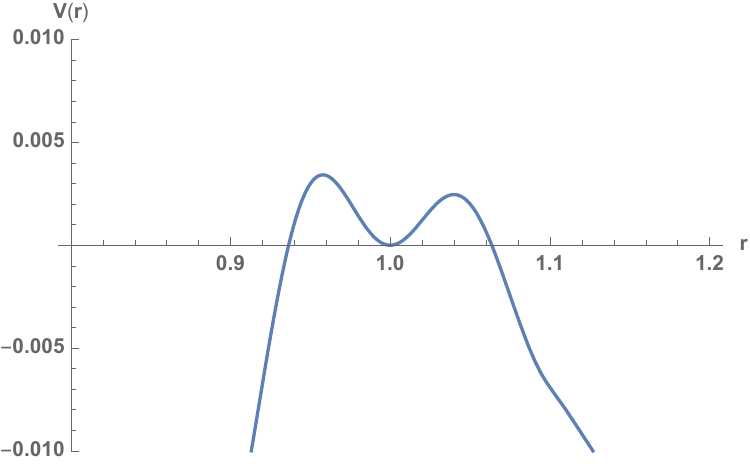}} \caption{ (colour online) The potential $V(r)=-\frac{1}{2\sigma^2 r^4}(1+\tilde V(r))$ with $1+\tilde V$ as in Eqn.\eqref{eq1pvtilde} with a perturbation as shown in Fig.\eqref{1pvtilde}. Note here that there is a local minimum at $r=1$. The parameters used are $\Lambda=0.1$, $M=1$, $\sigma=2$, $ r_*=0.3$ and $\rho=0.05$.}\label{potp}
\end{figure}

\subsection{Stable, static solutions that satisfy the dominant energy condition everywhere}\label{subsection5}
\begin{figure}[H]
\centerline{\includegraphics[width=0.5\linewidth]{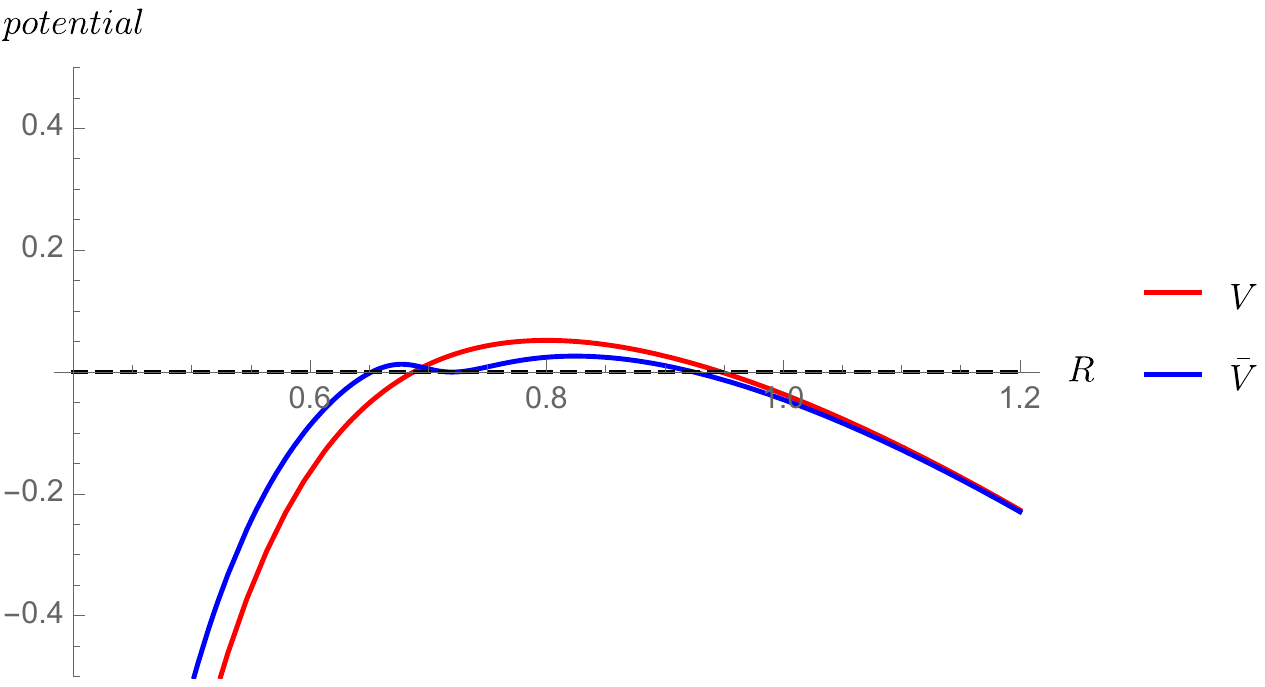}} \caption{(colour online) The potential $V(r)$ for constant $\sigma=.04$  (red) and the modified potential $\bar V(r)$ (blue) with $\sigma(r)$ given by Eqn.\eqref{sigmar}, and $M=.045, \Lambda_i/\Lambda=.994$.}\label{VVtilde}
\end{figure}

We can find stable static thin wall  solutions that everywhere satisfy the dominant energy condition, including on the wall, if we allow the energy density on the wall to depend on the radius of the bubble.  Such energy momentum on the wall, which knows about the radius of the bubble because it gives rise to the curvature of the lower dimensional space-time in the wall, is somewhat non-standard however it is perfectly physical.   Thus if we generalize $\sigma\to \sigma(r)$ in Eqn.\eqref{radius} through Eqn.\eqref{21} we find, with $\sigma\to (\sqrt{\Lambda/3}/4\pi)\sigma$ , $r\to \sqrt{3/\Lambda}\, \, r$ and $M\to (1/\sqrt{3\Lambda})M$,
\beq
V(r)=-\frac{M^2}{18 \sigma^2 r^4} +\frac{M \left((1-{\lb\Lambda_i/\Lambda\rb})+\sigma^2\right)}{6 \sigma^2 r}-\frac{r^2 \left(2 {\lb\Lambda_i/\Lambda\rb} \sigma^2+(1-{\lb\Lambda_i/\Lambda\rb})^2+\sigma^4+2 \sigma^2\right)}{8 \sigma^2}+\haf 1.
\eeq
Then we make $\sigma$  a function of $r$ to obtain a potential with a stable minimum of the required type.  If we leave $\sigma$ a constant, we obtain the potential like the example given by the graph in Fig.\eqref{fig0}.  Now we modify, using simple numerical experimentation.  If initially $\sigma=.04$ and then we modify it as
\beq
\sigma(r)= .04 - .0035 \tanh\lb\frac{25}{r_0} (r - r_0 + .105)\rb\label{sigmar}
\eeq
where $r_0$ is the position of the maximum of the potential when $\sigma=.04$ is a constant, (in Eqn.\eqref{r0} we had also imposed that $V(r_0)=0$, but here we do not, since it is only the modified potential that must satisfy this condition) we find that the potential has the required stable minimum as shown in Fig.\eqref{VVtilde}.  Here the potential $V(r)$ for constant $\sigma$ is shown and the modified potential with the stable minimum is called $\bar V(r)$.

\begin{figure}[H]
\centerline{\includegraphics[width=0.5\linewidth]{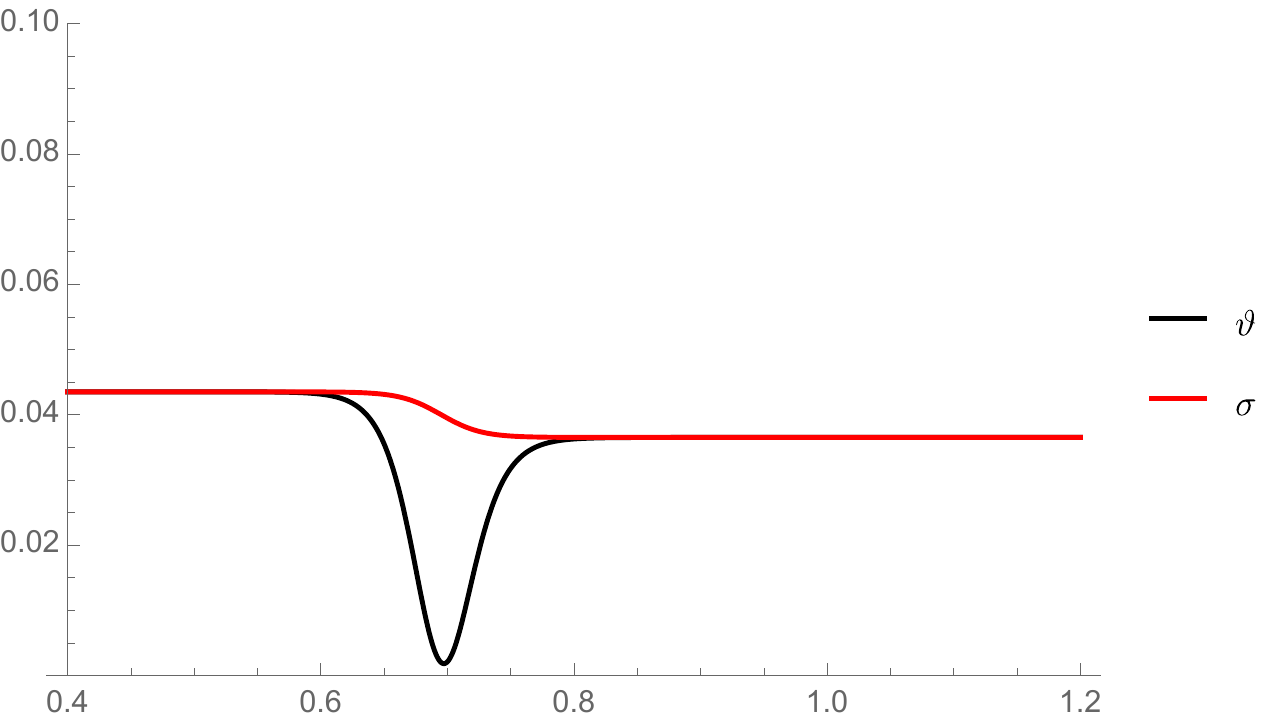}} \caption{(colour online) $\sigma(r)$ (red) given by Eqn.\eqref{sigmar} and $\vartheta(r)$ (black)given by Eqn.\eqref{vartheta} and $M=.045, \Lambda_i/\Lambda=.994$. }\label{sigmavartheta}
\end{figure}

The dominant energy condition is evidently satisfied in the bulk, as the metric describes a pure de Sitter space-time inside and a negative mass Schwarzschild-de Sitter space-time on the outside (both of which satisfy the dominant energy condition).   On the wall, the spatial stress \cite{Visser:2003ge} is given by
\beq
\vartheta(r)=\sigma(r)+\frac{r}{2}\frac{d\sigma(r)}{dr}\label{vartheta}
\eeq
and we also find numerically that the dominant energy condition on the wall, $\sigma\ge |\vartheta|$, is satisfied as can be seen in Fig.\eqref{sigmavartheta}.
As our goal was to provide a `proof-of-principle', the particular form of $\sigma(r)$ is not terribly important, just that there exists a configuration that respects the dominant energy condition. Although such a thin-wall configuration is not the limit of a single scalar field coupled to gravity, it is plausibly the limit of a theory with multiple scalar fields. In such a theory, one scalar provides the double-well potential, and the other scalars can be designed with couplings that are only active inside the wall, yielding a dynamical energy density on the shell. An example of this phenomenon can be found in \cite{Aguirre:2009tp}; although this particular example is not relevant to finding negative mass solutions, it illustrates that multi-scalar theories can give rise to dynamics on the wall. 

\subsection{Conclusions}   
We have shown that there exist stable (against spherical perturbations), static  negative mass bubble type solutions in the Schwarzschild-de Sitter space-time.  The bubble wall is considered to be thin, and the Israel junction conditions are imposed across the wall.  The junction conditions can be interpreted as giving the inside mass parameter $m_-(r)$ as a functional of a potential $V(r)$ that the radius of the bubble wall is sensitive to.  In this paper we have shown, under the assumption of spherical symmetry and with constant positive energy density $\sigma\ge0$ on the wall,  it is not possible to choose this potential in such a way that the bubble wall finds itself at a stable, classical minimum of the potential while determining an interior mass function in such a way that it is non-singular and that the dominant energy condition is everywhere satisfied.  If we allow $\sigma\le 0$ then the bubble wall  carries all the negative mass and does not respect the dominant energy condition, however stable solutions do exist and the dominant energy conditions can be made to be respected in the bulk away from the wall.  It is as if the negative mass is localized on the bubble wall.  We then allow the energy density on the wall to be non-constant but positive, $\sigma(r)\ge 0$.  Then with the simple ansatz of a pure de Sitter space-time in the interior and a negative mass Schwarzschild-de Sitter space-time in the exterior, we can find solutions corresponding to stable, non-singular, negative mass bubbles that respect the dominant energy condition everywhere including on the wall.   The formation of such configurations in the early universe must have important consequences for its subsequent evolution.

We have further demonstrated that there exist perfectly physical initial conditions for the matter distribution which can collapse to a singular, negative mass Schwarzschild-de Sitter space-time.  The initial conditions correspond to a bubble in negative mass Schwarzschild-de Sitter space-time with a bubble with the interior given by a simple de Sitter space-time separated by a thin wall.   The matter distribution actually satisfies the dominant energy condition everywhere including on the bubble wall.    The bubble can collapse through spherically symmetric configurations, to the exact negative mass Schwarzschild-de Sitter metric, which is singular at the origin and is not hidden behind an event horizon.  This is a counter example to the cosmic censorship hypothesis.

\subsection{Acknowledgments} We thank Emil Mottola and Edward Wilson-Ewing for useful discussions.  We thank NSERC of Canada for financial support and The Perimeter Institute for Theoretical Physics for hospitality.  N. T. thanks to the Conicyt scholarship 21160064 and the University of Santiago de Chile. Research at Perimeter Institute is supported by the Government of Canada through
the Department of Innovation, Science and Economic Development Canada and by the Province of Ontario
through the Ministry of Research, Innovation and Science. 

\bibliographystyle{apsrev}
\bibliography{ref1}

\begin{thebibliography}{32}
\expandafter\ifx\csname natexlab\endcsname\relax\def\natexlab#1{#1}\fi
\expandafter\ifx\csname bibnamefont\endcsname\relax
  \def\bibnamefont#1{#1}\fi
\expandafter\ifx\csname bibfnamefont\endcsname\relax
  \def\bibfnamefont#1{#1}\fi
\expandafter\ifx\csname citenamefont\endcsname\relax
  \def\citenamefont#1{#1}\fi
\expandafter\ifx\csname url\endcsname\relax
  \def\url#1{\texttt{#1}}\fi
\expandafter\ifx\csname urlprefix\endcsname\relax\def\urlprefix{URL }\fi
\providecommand{\bibinfo}[2]{#2}
\providecommand{\eprint}[2][]{\url{#2}}

\bibitem[{\citenamefont{Bellet{\^e}te and Paranjape}(2013)}]{Belletete:2013nqa}
\bibinfo{author}{\bibfnamefont{J.}~\bibnamefont{Bellet{\^e}te}}
  \bibnamefont{and} \bibinfo{author}{\bibfnamefont{M.~B.}
  \bibnamefont{Paranjape}}, \bibinfo{journal}{Int. J. Mod. Phys.}
  \textbf{\bibinfo{volume}{D22}}, \bibinfo{pages}{1341017}
  (\bibinfo{year}{2013}), \eprint{1304.1566}.

\bibitem[{\citenamefont{Mbarek and Paranjape}(2014)}]{Mbarek:2014ppa}
\bibinfo{author}{\bibfnamefont{S.}~\bibnamefont{Mbarek}} \bibnamefont{and}
  \bibinfo{author}{\bibfnamefont{M.~B.} \bibnamefont{Paranjape}},
  \bibinfo{journal}{Phys. Rev.} \textbf{\bibinfo{volume}{D90}},
  \bibinfo{pages}{101502} (\bibinfo{year}{2014}), \eprint{1407.1457}.

\bibitem[{\citenamefont{Schon and Yau}(1979)}]{Schon:1979rg}
\bibinfo{author}{\bibfnamefont{R.}~\bibnamefont{Schon}} \bibnamefont{and}
  \bibinfo{author}{\bibfnamefont{S.-T.} \bibnamefont{Yau}},
  \bibinfo{journal}{Commun. Math. Phys.} \textbf{\bibinfo{volume}{65}},
  \bibinfo{pages}{45} (\bibinfo{year}{1979}).

\bibitem[{\citenamefont{Schon and Yau}(1981)}]{Schon:1981vd}
\bibinfo{author}{\bibfnamefont{R.}~\bibnamefont{Schon}} \bibnamefont{and}
  \bibinfo{author}{\bibfnamefont{S.-T.} \bibnamefont{Yau}},
  \bibinfo{journal}{Commun. Math. Phys.} \textbf{\bibinfo{volume}{79}},
  \bibinfo{pages}{231} (\bibinfo{year}{1981}).

\bibitem[{\citenamefont{Witten}(1981)}]{Witten:1981mf}
\bibinfo{author}{\bibfnamefont{E.}~\bibnamefont{Witten}},
  \bibinfo{journal}{Commun. Math. Phys.} \textbf{\bibinfo{volume}{80}},
  \bibinfo{pages}{381} (\bibinfo{year}{1981}).

\bibitem[{\citenamefont{Arnowitt et~al.}(2008)\citenamefont{Arnowitt, Deser,
  and Misner}}]{Arnowitt:1962hi}
\bibinfo{author}{\bibfnamefont{R.~L.} \bibnamefont{Arnowitt}},
  \bibinfo{author}{\bibfnamefont{S.}~\bibnamefont{Deser}}, \bibnamefont{and}
  \bibinfo{author}{\bibfnamefont{C.~W.} \bibnamefont{Misner}},
  \bibinfo{journal}{Gen. Rel. Grav.} \textbf{\bibinfo{volume}{40}},
  \bibinfo{pages}{1997} (\bibinfo{year}{2008}), \eprint{gr-qc/0405109}.

\bibitem[{\citenamefont{Mann}(1997)}]{Mann:1997jb}
\bibinfo{author}{\bibfnamefont{R.~B.} \bibnamefont{Mann}},
  \bibinfo{journal}{Class. Quant. Grav.} \textbf{\bibinfo{volume}{14}},
  \bibinfo{pages}{2927} (\bibinfo{year}{1997}), \eprint{gr-qc/9705007}.

\bibitem[{\citenamefont{Smith and Mann}(1997)}]{Smith:1997wx}
\bibinfo{author}{\bibfnamefont{W.~L.} \bibnamefont{Smith}} \bibnamefont{and}
  \bibinfo{author}{\bibfnamefont{R.~B.} \bibnamefont{Mann}},
  \bibinfo{journal}{Phys. Rev.} \textbf{\bibinfo{volume}{D56}},
  \bibinfo{pages}{4942} (\bibinfo{year}{1997}), \eprint{gr-qc/9703007}.

\bibitem[{\citenamefont{Ashtekar
  et~al.}(2015{\natexlab{a}})\citenamefont{Ashtekar, Bonga, and
  Kesavan}}]{Ashtekar:2014zfa}
\bibinfo{author}{\bibfnamefont{A.}~\bibnamefont{Ashtekar}},
  \bibinfo{author}{\bibfnamefont{B.}~\bibnamefont{Bonga}}, \bibnamefont{and}
  \bibinfo{author}{\bibfnamefont{A.}~\bibnamefont{Kesavan}},
  \bibinfo{journal}{Class. Quant. Grav.} \textbf{\bibinfo{volume}{32}},
  \bibinfo{pages}{025004} (\bibinfo{year}{2015}{\natexlab{a}}),
  \eprint{1409.3816}.

\bibitem[{\citenamefont{Ashtekar
  et~al.}(2015{\natexlab{b}})\citenamefont{Ashtekar, Bonga, and
  Kesavan}}]{Ashtekar:2015lla}
\bibinfo{author}{\bibfnamefont{A.}~\bibnamefont{Ashtekar}},
  \bibinfo{author}{\bibfnamefont{B.}~\bibnamefont{Bonga}}, \bibnamefont{and}
  \bibinfo{author}{\bibfnamefont{A.}~\bibnamefont{Kesavan}},
  \bibinfo{journal}{Phys. Rev.} \textbf{\bibinfo{volume}{D92}},
  \bibinfo{pages}{044011} (\bibinfo{year}{2015}{\natexlab{b}}),
  \eprint{1506.06152}.

\bibitem[{\citenamefont{Ashtekar
  et~al.}(2015{\natexlab{c}})\citenamefont{Ashtekar, Bonga, and
  Kesavan}}]{Ashtekar:2015lxa}
\bibinfo{author}{\bibfnamefont{A.}~\bibnamefont{Ashtekar}},
  \bibinfo{author}{\bibfnamefont{B.}~\bibnamefont{Bonga}}, \bibnamefont{and}
  \bibinfo{author}{\bibfnamefont{A.}~\bibnamefont{Kesavan}},
  \bibinfo{journal}{Phys. Rev.} \textbf{\bibinfo{volume}{D92}},
  \bibinfo{pages}{104032} (\bibinfo{year}{2015}{\natexlab{c}}),
  \eprint{1510.05593}.

\bibitem[{\citenamefont{Israel}(1966)}]{Israel:1966rt}
\bibinfo{author}{\bibfnamefont{W.}~\bibnamefont{Israel}},
  \bibinfo{journal}{Nuovo Cim.} \textbf{\bibinfo{volume}{B44S10}},
  \bibinfo{pages}{1} (\bibinfo{year}{1966}), \bibinfo{note}{[Nuovo
  Cim.B44,1(1966)]}.

\bibitem[{\citenamefont{Visser and Wiltshire}(2004)}]{Visser:2003ge}
\bibinfo{author}{\bibfnamefont{M.}~\bibnamefont{Visser}} \bibnamefont{and}
  \bibinfo{author}{\bibfnamefont{D.~L.} \bibnamefont{Wiltshire}},
  \bibinfo{journal}{Class. Quant. Grav.} \textbf{\bibinfo{volume}{21}},
  \bibinfo{pages}{1135} (\bibinfo{year}{2004}), \eprint{gr-qc/0310107}.

\bibitem[{\citenamefont{Berezin et~al.}(1983)\citenamefont{Berezin, Kuzmin, and
  Tkachev}}]{Berezin:1982ur}
\bibinfo{author}{\bibfnamefont{V.~A.} \bibnamefont{Berezin}},
  \bibinfo{author}{\bibfnamefont{V.~A.} \bibnamefont{Kuzmin}},
  \bibnamefont{and} \bibinfo{author}{\bibfnamefont{I.~I.}
  \bibnamefont{Tkachev}}, \bibinfo{journal}{Phys. Lett.}
  \textbf{\bibinfo{volume}{120B}}, \bibinfo{pages}{91} (\bibinfo{year}{1983}).

\bibitem[{\citenamefont{Blau et~al.}(1987)\citenamefont{Blau, Guendelman, and
  Guth}}]{Blau:1986cw}
\bibinfo{author}{\bibfnamefont{S.~K.} \bibnamefont{Blau}},
  \bibinfo{author}{\bibfnamefont{E.~I.} \bibnamefont{Guendelman}},
  \bibnamefont{and} \bibinfo{author}{\bibfnamefont{A.~H.} \bibnamefont{Guth}},
  \bibinfo{journal}{Phys. Rev.} \textbf{\bibinfo{volume}{D35}},
  \bibinfo{pages}{1747} (\bibinfo{year}{1987}).

\bibitem[{\citenamefont{Farhi and Guth}(1987)}]{Farhi:1986ty}
\bibinfo{author}{\bibfnamefont{E.}~\bibnamefont{Farhi}} \bibnamefont{and}
  \bibinfo{author}{\bibfnamefont{A.~H.} \bibnamefont{Guth}},
  \bibinfo{journal}{Phys. Lett.} \textbf{\bibinfo{volume}{B183}},
  \bibinfo{pages}{149} (\bibinfo{year}{1987}).

\bibitem[{\citenamefont{Berezin et~al.}(1987)\citenamefont{Berezin, Kuzmin, and
  Tkachev}}]{Berezin:1987bc}
\bibinfo{author}{\bibfnamefont{V.~A.} \bibnamefont{Berezin}},
  \bibinfo{author}{\bibfnamefont{V.~A.} \bibnamefont{Kuzmin}},
  \bibnamefont{and} \bibinfo{author}{\bibfnamefont{I.~I.}
  \bibnamefont{Tkachev}}, \bibinfo{journal}{Phys. Rev.}
  \textbf{\bibinfo{volume}{D36}}, \bibinfo{pages}{2919} (\bibinfo{year}{1987}).

\bibitem[{\citenamefont{Farhi et~al.}(1990)\citenamefont{Farhi, Guth, and
  Guven}}]{Farhi:1989yr}
\bibinfo{author}{\bibfnamefont{E.}~\bibnamefont{Farhi}},
  \bibinfo{author}{\bibfnamefont{A.~H.} \bibnamefont{Guth}}, \bibnamefont{and}
  \bibinfo{author}{\bibfnamefont{J.}~\bibnamefont{Guven}},
  \bibinfo{journal}{Nucl. Phys.} \textbf{\bibinfo{volume}{B339}},
  \bibinfo{pages}{417} (\bibinfo{year}{1990}).

\bibitem[{\citenamefont{Aurilia et~al.}(1989)\citenamefont{Aurilia, Palmer, and
  Spallucci}}]{Aurilia:1989sb}
\bibinfo{author}{\bibfnamefont{A.}~\bibnamefont{Aurilia}},
  \bibinfo{author}{\bibfnamefont{M.}~\bibnamefont{Palmer}}, \bibnamefont{and}
  \bibinfo{author}{\bibfnamefont{E.}~\bibnamefont{Spallucci}},
  \bibinfo{journal}{Phys. Rev.} \textbf{\bibinfo{volume}{D40}},
  \bibinfo{pages}{2511} (\bibinfo{year}{1989}).

\bibitem[{\citenamefont{Mazur and Mottola}(2004)}]{Mazur:2004fk}
\bibinfo{author}{\bibfnamefont{P.~O.} \bibnamefont{Mazur}} \bibnamefont{and}
  \bibinfo{author}{\bibfnamefont{E.}~\bibnamefont{Mottola}},
  \bibinfo{journal}{Proc. Nat. Acad. Sci.} \textbf{\bibinfo{volume}{101}},
  \bibinfo{pages}{9545} (\bibinfo{year}{2004}), \eprint{gr-qc/0407075}.

\bibitem[{\citenamefont{Aguirre and Johnson}(2005)}]{Aguirre:2005xs}
\bibinfo{author}{\bibfnamefont{A.}~\bibnamefont{Aguirre}} \bibnamefont{and}
  \bibinfo{author}{\bibfnamefont{M.~C.} \bibnamefont{Johnson}},
  \bibinfo{journal}{Phys. Rev.} \textbf{\bibinfo{volume}{D72}},
  \bibinfo{pages}{103525} (\bibinfo{year}{2005}), \eprint{gr-qc/0508093}.

\bibitem[{\citenamefont{Aguirre and Johnson}(2006)}]{Aguirre:2005nt}
\bibinfo{author}{\bibfnamefont{A.}~\bibnamefont{Aguirre}} \bibnamefont{and}
  \bibinfo{author}{\bibfnamefont{M.~C.} \bibnamefont{Johnson}},
  \bibinfo{journal}{Phys. Rev.} \textbf{\bibinfo{volume}{D73}},
  \bibinfo{pages}{123529} (\bibinfo{year}{2006}), \eprint{gr-qc/0512034}.

\bibitem[{\citenamefont{Barnaveli and Gogberashvili}(1997)}]{Barnaveli:1996pn}
\bibinfo{author}{\bibfnamefont{A.}~\bibnamefont{Barnaveli}} \bibnamefont{and}
  \bibinfo{author}{\bibfnamefont{M.}~\bibnamefont{Gogberashvili}},
  \bibinfo{journal}{Theor. Math. Phys.} \textbf{\bibinfo{volume}{113}},
  \bibinfo{pages}{1491} (\bibinfo{year}{1997}), \bibinfo{note}{[Teor. Mat.
  Fiz.113,346(1997)]}, \eprint{hep-ph/9610548}.

\bibitem[{\citenamefont{Barnaveli and Gogberashvili}(1995)}]{Barnaveli:1995ep}
\bibinfo{author}{\bibfnamefont{A.}~\bibnamefont{Barnaveli}} \bibnamefont{and}
  \bibinfo{author}{\bibfnamefont{M.}~\bibnamefont{Gogberashvili}}, pp.
  \bibinfo{pages}{5--44} (\bibinfo{year}{1995}), \eprint{hep-ph/9505412}.

\bibitem[{\citenamefont{Barnaveli and Gogberashvili}(1994)}]{Barnaveli1994}
\bibinfo{author}{\bibfnamefont{A.}~\bibnamefont{Barnaveli}} \bibnamefont{and}
  \bibinfo{author}{\bibfnamefont{M.}~\bibnamefont{Gogberashvili}},
  \bibinfo{journal}{General Relativity and Gravitation}
  \textbf{\bibinfo{volume}{26}}, \bibinfo{pages}{1117} (\bibinfo{year}{1994}),
  ISSN \bibinfo{issn}{1572-9532},
  \urlprefix\url{https://doi.org/10.1007/BF02108937}.

\bibitem[{\citenamefont{Garriga and Megevand}(2004)}]{Garriga:2004nm}
\bibinfo{author}{\bibfnamefont{J.}~\bibnamefont{Garriga}} \bibnamefont{and}
  \bibinfo{author}{\bibfnamefont{A.}~\bibnamefont{Megevand}},
  \bibinfo{journal}{Int. J. Theor. Phys.} \textbf{\bibinfo{volume}{43}},
  \bibinfo{pages}{883} (\bibinfo{year}{2004}), \eprint{hep-th/0404097}.

\bibitem[{\citenamefont{Callan and Coleman}(1977)}]{Callan:1977pt}
\bibinfo{author}{\bibfnamefont{C.~G.} \bibnamefont{Callan}, \bibfnamefont{Jr.}}
  \bibnamefont{and} \bibinfo{author}{\bibfnamefont{S.~R.}
  \bibnamefont{Coleman}}, \bibinfo{journal}{Phys. Rev.}
  \textbf{\bibinfo{volume}{D16}}, \bibinfo{pages}{1762} (\bibinfo{year}{1977}).

\bibitem[{\citenamefont{Coleman}(1977)}]{Coleman:1977py}
\bibinfo{author}{\bibfnamefont{S.~R.} \bibnamefont{Coleman}},
  \bibinfo{journal}{Phys. Rev.} \textbf{\bibinfo{volume}{D15}},
  \bibinfo{pages}{2929} (\bibinfo{year}{1977}), \bibinfo{note}{[Erratum: Phys.
  Rev.D16,1248(1977)]}.

\bibitem[{\citenamefont{Coleman and De~Luccia}(1980)}]{Coleman:1980aw}
\bibinfo{author}{\bibfnamefont{S.~R.} \bibnamefont{Coleman}} \bibnamefont{and}
  \bibinfo{author}{\bibfnamefont{F.}~\bibnamefont{De~Luccia}},
  \bibinfo{journal}{Phys. Rev.} \textbf{\bibinfo{volume}{D21}},
  \bibinfo{pages}{3305} (\bibinfo{year}{1980}).

\bibitem[{\citenamefont{Penrose}(1969)}]{Penrose:1969pc}
\bibinfo{author}{\bibfnamefont{R.}~\bibnamefont{Penrose}},
  \bibinfo{journal}{Riv. Nuovo Cim.} \textbf{\bibinfo{volume}{1}},
  \bibinfo{pages}{252} (\bibinfo{year}{1969}), \bibinfo{note}{[Gen. Rel.
  Grav.34,1141(2002)]}.

\bibitem[{\citenamefont{Adams et~al.}(1990)\citenamefont{Adams, Freese, and
  Widrow}}]{Adams:1989su}
\bibinfo{author}{\bibfnamefont{F.~C.} \bibnamefont{Adams}},
  \bibinfo{author}{\bibfnamefont{K.}~\bibnamefont{Freese}}, \bibnamefont{and}
  \bibinfo{author}{\bibfnamefont{L.~M.} \bibnamefont{Widrow}},
  \bibinfo{journal}{Phys. Rev.} \textbf{\bibinfo{volume}{D41}},
  \bibinfo{pages}{347} (\bibinfo{year}{1990}).

\bibitem[{\citenamefont{Aguirre et~al.}(2010)\citenamefont{Aguirre, Johnson,
  and Larfors}}]{Aguirre:2009tp}
\bibinfo{author}{\bibfnamefont{A.}~\bibnamefont{Aguirre}},
  \bibinfo{author}{\bibfnamefont{M.~C.} \bibnamefont{Johnson}},
  \bibnamefont{and} \bibinfo{author}{\bibfnamefont{M.}~\bibnamefont{Larfors}},
  \bibinfo{journal}{Phys. Rev.} \textbf{\bibinfo{volume}{D81}},
  \bibinfo{pages}{043527} (\bibinfo{year}{2010}), \eprint{0911.4342}.

\end{thebibliography}

\end{document}